\documentclass{article}
\usepackage[dvips]{graphicx}

\newcommand\beq{\begin{equation}}
\newcommand\eeq{\end{equation}}
\newcommand\bea{\begin{eqnarray}}
\newcommand\eea{\end{eqnarray}}
\newcommand\FF{{\cal F}}

\newcommand\bdm{\begin{displaymath}}
\newcommand\edm{\end{displaymath}}

\begin{document}

\begin{flushright}
\vskip-0.75cm
UdeM-GPP-TH-02-95\\
\end{flushright}
\bigskip

\begin{center}
{\LARGE Magnetic properties of SO(5) superconductivity}\\
\bigskip\bigskip
M.\ Juneau, R.\ MacKenzie, M.-A.\ Vachon\\
\it Laboratoire Ren\'e-J.-A.-L\'evesque, Universit\'e de Montr\'eal\\
C.P. 6128, Succ. Centre-ville, Montr\'eal, QC H4V 2A5
\end{center}

\bigskip
\begin{abstract}
The distinction between type I and type II superconductivity is
re-examined in the context of the SO(5) model recently put forth by
Zhang. Whereas in conventional superconductivity only one parameter
(the Ginz\-burg-Landau parameter $\kappa$)
characterizes the model, in the SO(5) model there are two essential
parameters. These can be chosen to be $\kappa$ and another parameter,
$\beta$, related to the doping.
There is a more complicated relation between
$\kappa$ and the behavior of a
superconductor in a magnetic field. In particular, one can find type I 
superconductivity, even when $\kappa$ is large, for appropriate values 
of $\beta$.

\end{abstract}
\bigskip

\section{Introduction}

Several years ago, the SO(5) model was proposed as a description of
high-temperature superconductors \cite{Zhang}. At low temperatures 
these materials exhibit superconductivity (SC) or antiferromagnetism 
(AF) depending on the doping; the SO(5) model attempts to unify these 
two phenomena (both of which involve spontaneous symmetry breaking) 
into a single symmetry group.

Not surprisingly, the SO(5) model is somewhat more complicated than
the corresponding model describing conventional superconductivity
(which is based on the smaller group SO(2)); correspondingly, richer
behavior can be seen.  In conventional superconductivity, one
parameter, the Ginzburg-Landau (GL) parameter $\kappa$, determines the
behavior. For instance, $\kappa=1/\sqrt2\equiv\kappa_c$ marks the
boundary between type I and type II superconductors.

In the SO(5) model, there are two dimensionless parameters, which can
be taken to be $\kappa$ and a second parameter, $\beta$ (to be defined
below); the latter is related to the degree to which SO(5) is explicitly
broken (by doping, for instance).  As will be shown below, this second
parameter has a profound influence on the magnetic behavior of
an SO(5) superconductor; for instance, the critical value of $\kappa$ in
the SO(5) model is given by
$\kappa_c(\beta)=(1/\sqrt2)\sqrt{(1+\beta)/(1-\beta)}$. One sees that as
$\beta\to1^-$, $\kappa_c\to\infty$, which is a dramatic departure from the
conventional value. This is of some significance since
high-temperature superconductors typically have $\kappa\gg1$, 
and therefore are normally thought to be extreme type II 
superconductors. Such a
conclusion would be premature in the SO(5) model, however, since (as
the above relation shows) one could have $\kappa_c>\kappa\gg1$.

In this paper we will first remind the reader of the magnetic
properties of conventional superconductors. The standard approach
based on the surface energy density of a boundary between a normal
phase at critical magnetic field $H_c$ and a superconducting phase
will be briefly reviewed. We will then discuss an
alternative approach based on vortex energetics.
Although much of the above is fairly familiar material, this
discussion will establish notation and set the stage for the parallel
discussion in the context of the SO(5) model.

Next, we will discuss the case of SO(5) superconductivity.  After a
brief review of the model itself, the critical fields $H_c$ and
$H_{c2}$ will be calculated. From these, the critical value of the
parameter $\kappa$ can be calculated explicitly as a function of
$\beta$. This expression can be confirmed numerically, either by
examining the surface energy density or vortex energetics. In the
latter approach, one can see that the dramatically different behavior
in the SO(5) model is due to the possibility of an antiferromagnetic
core in the vortex \cite{Arovas,Alama,Mack}.

\section[Review]{Review of conventional superconductivity\footnote{For a 
more complete discussion on conventional SC, see \cite{Saint-James} or 
\cite{Tinkham}}}\label{sc}

A conventional superconductor is described by the following Helmholtz free 
energy in the GL theory:
\beq\label{GLFE}
\hat{F}=\int{d\mathbf{x}} \left\{ f_n - \frac{a^2_1}{2} |\phi|^2 + \frac{b^2}
{4} |\phi|^4 + \frac{1}{2m^*} \left|\left(-i\hbar\nabla - \frac{e^*\mathbf{A}}
{c}\right)\phi\right|+\frac{\mathbf{h}^2}{8\pi}\right\},
\eeq
where $f_n$ is a constant, $\hat{\mathbf{h}}= \nabla \times \mathbf{A}$ is 
the microscopic magnetic field and $a_1$, $b$ are parameters. The minimum 
of the potential is $|\phi|^2 = a^2_1/b \equiv v^2$.

The Helmholtz free energy is minimized when the temperature $T$ and interior 
magnetic induction $B$ are constant (i.e., thermodynamic equilibrium). One 
uses the Gibbs free energy for equilibrium where $T$ and the external 
magnetic field $H$ are constant. These two quantities are related by the 
following Legendre transformation:\
\bea \label{legendre}
g(H,T) = f(h,T) - \frac{1}{4\pi}hH.
\eea
The second term in this equation is the microscopic version of 
$(4\pi)^{-1}BH$, since the magnetic induction $B(\mathbf{r})= \langle 
h(\mathbf{r}) \rangle$.

The critical magnetic field $H_c$ is defined as the field which establishes 
the following condition on the Gibbs energies in the normal and 
superconductor regions:
\bea\label{criticalc}
g_n(H_c,T)=g_s(H_c,T).
\eea
We also have the following relations, the first one due to the Meissner effect:
\bea
g_s(H,T)=f_s(T), \nonumber \\
g_n(H,T)=f_n - \frac{H^2}{4\pi}. \nonumber
\eea
From Eq.~(\ref{GLFE}), we can obtain a correspondence
with the parameters of the theory:
\bea
f_n(T)-f_s(T)=\frac{a^4_1}{4b}. \nonumber
\eea
One finds:
\bea
\frac{H^2_c}{4\pi}=\frac{a^4_1}{2b}
=\frac{\Phi^2_0}{32\pi ^3 \lambda ^2 \xi ^2},
\label{Hc}
\eea
where $\Phi_0=hc/e^*$ is the quantum of magnetic flux, $\lambda = 
(m^* c^2 / 4\pi e^{*^{2}} v^2)^{1/2}$ is the penetration depth and 
$\xi = (\hbar ^2 / m^* a^2_1)^{1/2}$ is the coherence length.

There is also another critical field, $H_{c2}$, which is the strongest field 
allowed in a superconducting region before the system goes to the normal 
state. It is given by:
\bea
H_{c2}=\frac{\Phi_0}{2\pi \xi^2}=\sqrt{2} \kappa H_c, \nonumber
\eea
where, as usual, $\kappa=\lambda/\xi$. We see that if $\kappa$ is equal 
to its critical value $1/\sqrt{2}$, $H_c=H_{c2}$ (more on this in Sections 
\ref{seconv} and \ref{hso5}). 

\subsection{Surface Energy}\label{seconv}

As is well known, a superconductor will behave in one of two ways when 
placed in a magnetic field. In a type I SC, macroscopic normal regions 
where $h=H_c$ will form, while in a type II SC, a lattice of vortices 
of flux $\Phi_0$ will appear. One way to determine under which circumstances 
a SC is of type I or II is by studying the energy per unit area of a 
boundary between a normal region and a SC region. If this energy is 
positive, the system will tend to minimize the interface's surface area 
in order to lower this energy as much as possible, indicative of a 
type I SC. Similarly for negative surface energy, this surface area will 
be maximized, inducing type II behavior.

Analytically, the surface energy is given by:
\bea
\hat{G}_s=\frac{\Phi^2_0 \kappa^2}{32 \pi^3 \lambda^3}I 
= \frac{H^2_c \lambda}{4 \pi}I, \nonumber
\eea
where:
\bea
I=\int_{-\infty}^{\infty} {ds}\left\{ \frac{1}{2}(1-f^4) 
+ h^2 -\sqrt{2} h \right\}.
\label{notreamii}
\eea

Consider therefore a field configuration between the normal 
state for $x \rightarrow - \infty$ and the SC state for $x 
\rightarrow +\infty$: 
\begin{displaymath}
\phi=\phi(x),
 \quad \mathbf{A} = A(x)\hat{y}.
\end{displaymath}
Let us go to a description in terms of dimensionless variables, setting:
\beq
\label{dimless}
x = \lambda s, \quad \phi = vf, \quad A = \frac{c \hbar }{ e^* \xi }a,  
\quad \hat{h} = \frac{c\hbar}{e^{*}\xi \lambda}h.
\eeq
We obtain the following dimensionless free energy:
\beq\label{fsdv}
F=\int {ds}\left\{ f_n + \frac{1}{\kappa ^2} f'^{2} 
+ (a^2 + f^2 - 1)f^2 + h^2 \right\}.
\eeq
Minimizing F yields the following equations (the so-called 1D GL equations):
\beq
\frac{1}{\kappa ^2} \frac{d^2 f}{ds^2} + (1-a^2)f -f^3=0,
\label{motion1}
\eeq
\beq
\frac{d^2 a}{ds^2}-af^2=0.
\label{motion2}
\eeq

The sign of the surface energy can be determined qualitatively in the
limit of either small or large value of $\kappa$ in
Eq.~(\ref{motion1}), by looking at the length scales of $h$ and
$f$. By construction (see Eq.~(\ref{dimless})), the length scale of
$h$ is $l_h=1$, while that of $f$ turns out to be
$l_f=1/(\sqrt{2}\kappa))$.

\begin{figure}[ht]
\centering
\includegraphics[angle=270,width=7.5cm]{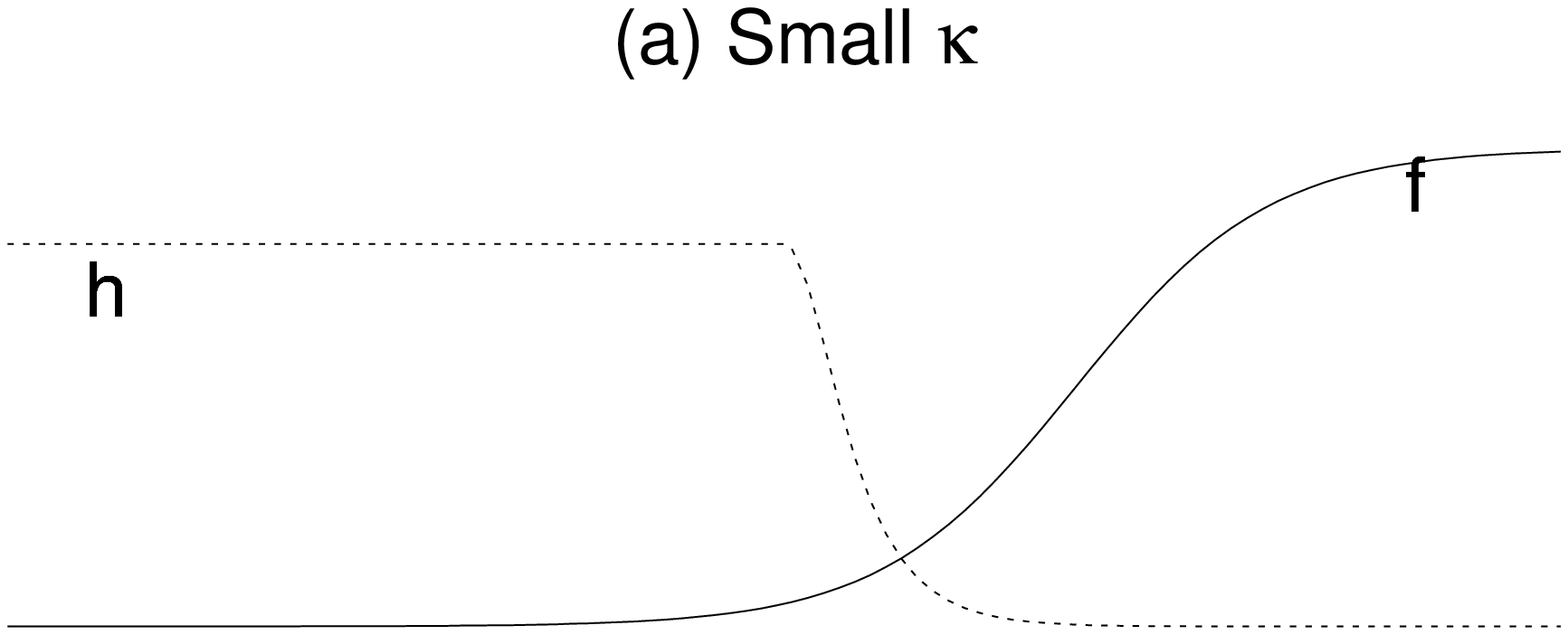}\\
\includegraphics[angle=270,width=7.5cm]{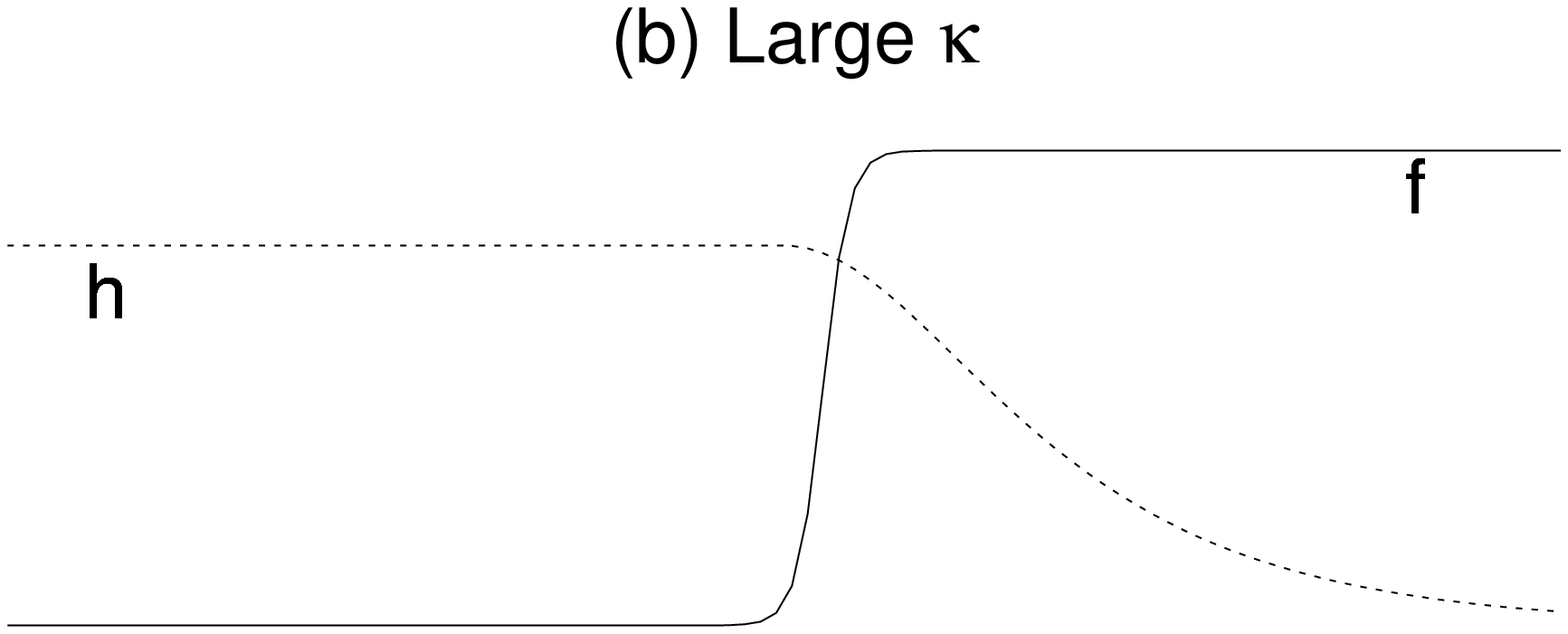}
\caption{Field solution behaviors for the conventional 
superconductor (a) type I (b) type II}
\label{charlen}
\end{figure}

From Fig.~\ref{charlen}, we can see how the sign of the surface energy
is determined by the behavior of $f$ and $h$ in those limiting
cases. In the first limiting case (Fig.~\ref{charlen}a) which
corresponds to a type I sample ($\kappa \ll 1$, so that $l_f \gg
l_h$), the energy comes from the expulsion of the magnetic field
within the SC region. More explicitly, $h$ is rapidly expulsed at the
interface while $f$ slowly assumes its asymptotic value. This gives
rise to a positive scalar field energy density without a cancelling
negative magnetic energy; the surface energy is
positive. Fig.~\ref{charlen}b shows the reverse case of a type II
sample, where $f$ quickly reaches its asymptotic behavior while the
magnetic field slowly decreases, resulting in a negative surface
energy.

The critical value of $\kappa$ is the one for which the surface energy
(or equivalently, the integral (\ref{notreamii})) is zero. The
well-known result $\kappa_c=1/\sqrt{2}$ can be determined either
numerically \footnote{ A numerical calculation of the surface energy is
given in \cite{Osborn}.}, or by making the oracular observation that
when $f$ and $h$ obey the equations of motion for $\kappa
=1/\sqrt{2}$, then the integrand of (\ref{notreamii}) is
zero\footnote{Note that there is no need for the integrand of
(\ref{notreamii}) itself to be zero; the weaker condition that the
integral is zero would suffice. Indeed, we will see in the next
section that in the SO(5) model the analogous integrand is \emph{not}
zero when the integrated surface energy is zero.}. This observation
can be explained by a recasting of the work of Bogomol'nyi
\cite{Bogol} (to be discussed in the next section) into a form
appropriate to the surface energy.
 
\subsection{Vortex Energetics}\label{veconv}

An alternative approach to determining the behavior of a
superconductor when placed in a magnetic field is the study of the
energy of vortices as a function of their winding number $m$. In this
section, we will see what information can be drawn from the analytic
expression, as well as an important limiting case, $m\to\infty$. In
this limit, we will also show how one can establish a link between the
vortex and surface energies.

As mentioned above, when placed in a magnetic field, the system will
eventually find the configuration which minimizes the free energy. We
can construct a function $\FF(m)=F(m)/m$ which is the free energy of a
vortex of winding number $m$ divided by the winding number, or in
other words the free energy per flux quantum of an $m$-vortex. For
large $m$, if $\FF(m)$ is an increasing function of $m$, it is more
energetically expensive for a magnetic field to penetrate in vortices
of large winding number; rather, a lattice of vortices of winding
number 1 will form. This, clearly, is what we expect of a type II
superconductor. Similarly, if $\FF(m)$ is a decreasing function of $m$,
the least energetic configuration will be a macroscopic normal region
containing a large number of flux quanta; this is what we expect of
a type I superconductor.

For the $m$-vortex configuration, we use an ansatz that is 
generalized to winding number $m$:
\begin{displaymath}
\phi=f(r) e^{im\theta}, \quad \hat{A_i} = \frac{a_1 c \sqrt{m^*}}{e^*} 
\epsilon_{ij} \frac{r_j}{r} a(r).
\end{displaymath}
Again we can use the dimensionless variables defined in
(\ref{dimless}). 
Here, however, $s$ denotes the radial variable in a 2D plane, 
as opposed to the cartesian coordinate orthogonal to the domain wall.

Using these changes in Eq.~(\ref{GLFE}) yields the vortex energy:
\begin{eqnarray}\label{fv}
F_v=\int_{0}^{\infty}{ds} \frac{s}{2} \bigg\{ \frac{1}{2} 
+ \left( a' + \frac{a}{s} \right) ^2 + \frac{1}{\kappa^2} 
\left( f'^{2} + \left ( \frac{m}{s} + \kappa a \right)^2 f^2 \right)
\nonumber\\
- (1 - f^2)f^2 \bigg\},
\end{eqnarray}
as well as the equations of motion:
\beq
\frac{1}{\kappa^2} \left( f'' + \frac{1}{s} f' 
- \left( \frac{m}{s} + \kappa a \right) ^2 f \right) + f - f^3 = 0,
\label{motion3}
\eeq
\beq
h' + \left( \frac{m}{\kappa s} + a \right) f^2 = 0.
\label{motion4}
\eeq

It is possible to rewrite Eq.~(\ref{fv}) in the following form \cite{Bogol}:
\bea\label{Bogv}
2\kappa ^2 F(m) = \int_{0}^{\infty} s {ds} \bigg\{ \left[ f' 
- \left( \frac{m}{s} + \kappa a \right)f \right] ^2
\nonumber\\
+ \left[ \kappa \left( a' + \frac{a}{s} \right) 
+ \frac{1}{2}(1-f^2)\right] ^2 + \frac{1}{2}(\kappa ^2 
- \frac{1}{2})(1-f^2) ^2 + \frac{d\chi}{ds}+\frac1s\chi \bigg\},
\eea
where $\chi = \kappa a (f^2-1) + mf^2/s$. Integrating the last two terms
yields the winding number $m$, independent of the details of the
field. Clearly, if $\kappa=1/\sqrt{2}$, the third term vanishes and
$F$ consists of the winding number $m$ plus two positive semidefinite
terms; it will be minimized if they are zero. One can see that these
terms are in fact zero if $f$ and $a$ satisfy the equations of
motion. In this case, $\FF(m)=F(m)/m=1$ independent of $m$, which
indeed corresponds to critical behavior with regard to vortice lattice
stability, as there is no preference for either type I (coalescing) or
type II (lattice) behavior.

We can determine the behavior of $\FF(m)$ numerically for a large
value of the winding number $m$. Our numerical results already suggest
that $\FF(m)$ reaches asymptotically a constant value at
$m\rightarrow\infty$. In this limiting case, the core of the vortex is
very wide, and the fields fluctuate rapidly at only one spatial
position-- that is at the vortex perimeter. Let us define this
position as $s_m$. Analytically, we will see this more clearly by
inserting the limiting forms of the field into Eq.~(\ref{Bogv}), which
are:
\bea
s < s_m :\qquad \qquad f \rightarrow 0, \qquad a \rightarrow 
-\frac{H_c}{2}s, \qquad h=0, \qquad \qquad \qquad \nonumber \\
s > s_m :\qquad \qquad f \rightarrow 1, \qquad a \rightarrow 
-\frac{m}{\kappa s}, \qquad h = -a' - \frac{a}{s} \rightarrow 
\frac{1}{\sqrt{2}}. \nonumber
\eea
After integration, the second and third terms of Eq.~(\ref{Bogv}) 
become, respectively:
\[
\left(\frac{1}{2} -\kappa H_c \right)^2 \frac{s_{m}^2}{2}
\qquad\mbox{and}\qquad
\frac{1}{4}(\kappa^2 -\frac{1}{2})s_{m}^2.
\]

We have $m$ and the remaining bracketed term, which does not
contribute to the integral (this, of course, stems from our choice of
step-functions as the role of fields $f$ and $h$, since we essentially
remove the presence of the surface altogether). Let us rewrite the
variable $s_m$ in terms of the quantum of magnetic flux $\Phi_0$,
using our $m \rightarrow \infty$ ans\"{a}tze:
\bea
\Phi_0 = \frac{1}{m} \int_{0}^{\infty} h{d^2x} = \frac{\pi s_m^2
H_c}{m} .
\nonumber
\eea
Therefore:
\bea
s_m^2=\frac{2m}{\kappa H_c}. \nonumber
\eea
We find:
\bea
\FF(m)=\frac{1}{\sqrt{2}\kappa} \qquad \qquad \qquad 
(m \rightarrow \infty). \nonumber
\eea

If we add thickness to the interface, the first term of
Eq.~(\ref{Bogv}), which was vanishing, will instead be proportional to
$\sqrt{m}$ (i.e., a ``perimeter'' term $\propto s_m$).

We are now able to establish a correspondence between the free
energies of the vortex and wall domain configurations. Indeed, for a
large value of $m$, the perimeter of the vortex is large enough to
significantly lower the curvature of the vortex locally, likening a
given small region into the form of a domain wall. We should expect to
see confirmation of this, using the appropriate approximations. We
will do this here.

Let us start from the Gibbs energy of the vortex and derive
an expression that clearly shows the relation with the domain wall
Gibbs energy. We have:
\[
\hat{G}_v=\hat{F}_v - \frac{H_c}{4\pi}\Phi_0 m
=\frac{\Phi_0 ^2}{16\pi^2 \lambda^2}\left( 2\kappa^2F(m) 
- \sqrt{2}\kappa m\right).
\]
Substituting $F(m)$ with Eq.~(\ref{Bogv}) and expanding the terms 
in square parentheses:
\bea
\hat{G}_v=\frac{\Phi^2_0}{16\pi^2 \lambda^2}
\left( m+ \int_{0}^{\infty} s{ds}\left\{ \frac{\kappa ^2}{2}
(1-f^2)^2 + \kappa^2 h^2 -\kappa h(1-f^2) \right. \right.
\nonumber \\
\label{MAV} \left. \left. + f'^{2} -2ff' 
\left( \frac{m}{s} + \kappa a\right) + 
\left( \frac{m}{s} + \kappa a \right) ^2f^2\right\} 
- \sqrt{2}\kappa m \right).
\eea

We notice that the following terms are equal to a familiar quantity: 
\bea
-\kappa h(1-f^2) - 2ff' \left( \frac{m}{s} + \kappa a\right) 
= -\frac{d\chi}{ds}, \nonumber
\eea
which, once integrated, cancels the $m$ term. 
If we take Eq.~(\ref{motion3}) and multiply it by $f\kappa ^2$, we have:
\bea
\left( \frac{m}{s} + \kappa a\right)^2 f^2=ff''+\kappa^2 (f^2 - f^4).
\label{subs}
\eea

Next, we change variables $s\to z=s-s_m$, where 
it is understood that the region of interest is $|z| \ll |s_m|$. Therefore
\bea
\int_{0}^{\infty} s\,{ds} \rightarrow \int_{-\infty}^{\infty}(z+s_m)\,{dz} 
\approx \int_{-\infty}^{\infty}s_m\,{dz}. \nonumber
\eea
Substituting (\ref{subs}) into (\ref{MAV}) and simplifying, we have:
\bea
\hat{G}_v=\frac{\kappa^2 \Phi_0^2}{16\pi^2 \lambda^2}
\left( \int_{-\infty}^{\infty}s_m{dz} \left\{ \frac{1}{2}(1-f^4) 
+ h^2 -\sqrt{2} h\right\} \right. \nonumber \\
\left. + \frac{s_m}{\kappa^2}\int_{-\infty}^{\infty}\frac{d(ff')}
{dz}{dz}\right).
\eea

We recognize the first integral: it is $s_m$ times $I$, 
as defined by Eq.~(\ref{notreamii}); the second integral 
is zero due to the boundary conditions. Therefore, the link 
between the Gibbs energies between the vortex and domain wall 
case is summed up by the following relation, valid for large $m$:
\bea
\hat{G}_v=\frac{\kappa^2 \Phi_0^2 s_m}{16\pi^2 \lambda^2}I
=\hat{G}_s (2\pi \lambda s_m). \nonumber
\eea

In this section, we have studied vortex energetics as an
alternative approach to the determination 
of the type (I vs. II) of a given superconductor, in the conventional
case. We now turn our attention to the case of SO(5) superconductivity. 

\section{SO(5) Superconductivity}

 Having briefly reviewed the basics of conventional SC, let us turn to
the case of SO(5) superconductivity.  As mentioned in the
introduction, the SO(5) model attempts to unify the superconducting and
antiferromagnetic aspects of high-temperature superconductors.  The 
order parameters for these two phenomena form a 5-dimensional real
field whose dynamics has an approximate SO(5) symmetry, $(\phi_1,
\phi_2,\eta_1, \eta_2,\eta_3)$, where $\phi=\phi_1 + i \phi_2$ and
$\mathbf{\eta}=( \eta_1, \eta_2,\eta_3)$ are, respectively, the order
parameters of superconductivity and antiferromagnetism.  This so-called
superspin is described by a free energy which will be given below; 
explicit SO(5) breaking enables spontaneous breaking of either
the SO(2) symmetry of $\phi$ or the SO(3) symmetry of $\mathbf{\eta}$,
describing SC and AF, respectively.

Mathematically, the low-energy effective theory is described in terms of
the following free energy \cite{Mack}:

\begin{equation}\label{free}
\hat{F} = \int d\mathbf{x} \left( \frac{ \hat{\mathbf{h}}^2}{8 \pi} 
+ \frac{ \hbar^2}{2 m^*}\left| \left( -i\nabla 
- \frac{e^*}{\hbar c}\hat{\mathbf{A}} \right) \phi \right|^2 
+ \frac{\hbar^2}{2m^*}( \nabla \mathbf{ \eta})^2 
+ V( \phi, \mathbf{\eta}) \right),
\end{equation}
where $ \hat{ \mathbf{h}} = \nabla \times \hat{\mathbf{A}}$ 
is again the microscopic magnetic field.

The nature of the ground state can be determined by
examining the potential.  We consider the most general
symmetry-breaking potential
including even powers of the fields up to quartic terms,
\begin{displaymath}\label{pot}
 V(\phi,\eta)= -\frac{a^2_1}{2} \phi^2 - \frac{a^2_2}{2} \eta^2 
+ \frac{ b_1 \phi^4 + 2 b_3 \phi^2 \eta^2 + b_2 \eta^4}{4},
\end{displaymath}
where we have written $ \phi = | \phi |$ and $\eta =|\mathbf{\eta}|$.
This potential is invariant under an $SO(3) \times
SO(2)$ symmetry.  SO(5) symmetry would be attained by setting
the two mass parameters and the three
quartic coupling constants to be equal; for an approximate symmetry
their values are approximately equal.  To simplify the analysis we will
set the quartic parameters to a common value $b_1 = b_2 = b_3 = b$;
explicit symmetry breaking in the potential is found only in the
quadratic terms.

Since we are interested in the magnetic properties of the
superconducting state, we restrict ourselves to parameters which
describe a SC ground state.  This is done by requiring that the global
minimum of $V$ be on the $\phi$ axis.  The ground state will then have
the value $(\phi, \eta) = (v,0)$, where $v \equiv a_1/\sqrt{b}$;
the global condition is fulfilled if $\beta \equiv a_2^2/a_1^2 <
1$.  Note that we recover SO(5) symmetry if $\beta = 1$ and,
therefore, this is the value corresponding to critical doping. For
$\beta>1$ the ground state is antiferromagnetic.

The goal of this section is to calculate the surface energy, whose
sign indicates the type of superconductor described by the model.
We will first set up the physical context of the surface energy and
derive the critical fields $H_c$ and $H_{c2}$. These
enable us to determine the value of $\kappa_c$ as a function of
$\beta$. Subsequently, we will give
numerical calculations which confirm our analytical results.

\subsection{One Dimensional Free Energy}\label{feso5}

As outlined in Section~\ref{seconv}, the surface energy is the Gibbs
free energy per unit area of a domain wall separating a normal
(non-superconducting) phase
in the thermodynamic critical field $H_c$ and a superconducting phase
in the absence of a magnetic field. 
We use the same ansatz for $\phi$ and $A$ as mentioned in the 
beginning of Section~\ref{seconv}, and with the AF order parameter 
written $ \eta = \eta(x) \hat{x}$, the free energy (\ref{free}) becomes:
\begin{eqnarray}\label{fs}
\hat{F} =\int{dx}\left\{  f_n + \frac{\hbar^2}{2m^*}
\left( \frac{d\phi}{dx}\right)^2 + \frac{1}{2m^*}
\left(\frac{e^*}{c}\right)^2A^2\phi^2 
+ \frac{1}{8\pi}\left(\frac{dA}{dx}\right)^2 \right. 
\nonumber\\
\left. + \frac{\hbar^2}{2m^*}\left(\frac{d\eta}{dx}\right)^2 
+ V(\phi,\eta) \right\}. 
\end{eqnarray}
It will again be useful to use the dimensionless quantities defined 
in (\ref{dimless}). Writing $\eta(x) = v n(s)$, we find, after 
some algebra, the following dimensionless free energy:
\begin{eqnarray}\label{fs1}
F= \int dx \left\{ f_n + \frac{1}{\kappa^2} (f'^2 + n'^2) 
+ a^2f^2 + h^2 -f^2 - \beta n^2 \right.
\nonumber\\
\left. + \frac{1}{2}( f^4 + 2f^2n^2 + n^4) \right\}.
\end{eqnarray}
Minimizing with respect to $f$, $n$ and $a$, we obtain the 
SO(5) GL equations: 
\begin{equation}\label{m1}
\frac{1}{\kappa^2}\frac{d^2f}{ds^2} 
+ \left(1 - a^2 \right)f - f^3 - n^2f=0,
\end{equation}
\begin{equation}\label{m2}
\frac{d^2a}{ds^2} = af^2,
\end{equation}
\begin{equation}\label{m3}
\frac{1}{\kappa^2}\frac{d^2n}{ds^2} +  \beta n - n^3 - nf^2 = 0.
\end{equation}
Note that if we set $n=0$ in the first two equations, we recover the
conventional GL Eqs.~(\ref{motion1}-\ref{motion2}), as expected.  Note
also that we are now left with two parameters ($\kappa$, $\beta$)
rather than three ($a_1^2$, $a_2^2$ and $b$).  The first of these two
is the traditional GL parameter $\kappa = \lambda/\xi$, which is
usually quite large for high $T_c$ SC (typically between 15 and
100).  The second fundamental parameter is $\beta$, which is
related to the doping of the system away from critical doping (that
corresponding to the AF/SC transition).  $\beta$ can be written as a
function of the chemical potential $ \mu$ as $ \beta = 1 - 8 m^*
\xi^2\chi( \mu^2 - \mu_c^2)/ \hbar^2$, where $\chi$ is the charge
susceptibility and $\mu_c$ is the critical value of the chemical
potential \cite{Arovas}.

\subsection{ Calculation of $H_c$ and $H_{c2}$}\label{hso5}

When a critical magnetic field is applied in a region, the SC order
parameter will be forced to zero, destroying superconductivity. In
order to minimize the Gibbs free energy, a nonzero AF order parameter
will be induced. This is because, $f$ being held to
zero by the magnetic field, the potential in this region is minimized
if $n = \sqrt{\beta}$, as can be seen from the potential terms in
(\ref{fs1}). The value of the critical field is determined by the
competition between the positive potential energy of the AF state
(relative to the SC ground state), on the one hand, and the negative
magnetic energy ($-H_c^2/8\pi$), on the other.

Let us compute, first of all, the critical magnetic field $H_c$. The result 
differs from the conventional one because an AF order parameter will now 
appear in the normal region. 

Far into the superconducting region ($ x \rightarrow \infty$),
we get the following behavior for the fields: $h=0$, $f=1$, $n=0$. Thus
from Eqs.~(\ref{legendre}) and (\ref{fs}),
\begin{equation}\label{gs}
 g_s(H,T) \approx g(h=0,T) = f_s(T) = f_n(T) - \frac{a_1^4}{4b}.
\end{equation}
In the normal region ($ x \rightarrow -\infty$),
we include a nonzero AF order parameter, and set $h=H$,
$n = \sqrt{\beta}$, $f = 0$, giving,
\begin{equation}\label{gn}
g_n(H,T) = f_n(T) - \frac{a_2^4}{4b} - \frac{H^2}{8 \pi}.
\end{equation}

By  equating the two Gibbs free energies, we find the following 
$\beta$-dependent expression for the critical magnetic field:
\begin{equation}\label{h1}
H_c(\beta)= H_c^0 \sqrt{1- \beta^2},
\end{equation} 
where $H_c^0$ is the conventional expression (\ref{Hc}).  
Note that $H_c(\beta)$ will refer to the critical field in the SO(5) model, 
and must be distinguished from the one in the GL theory (from now on, 
we will write $H_c^0$). 
Eq.~(\ref{h1}) tells us that for
$\beta=1$, $H_c(\beta)$ vanishes.  In fact, this value of $\beta$ corresponds 
to the critical value, where the transition from SC ($\beta < 1$) to AF
($\beta > 1$) occurs; it is natural that any magnetic field, however small, 
will break SC, since SC is not energetically favored over AF: their free 
energies are equal.
As $\beta$ decreases, $H_c(\beta)$ reaches a maximum when $\beta=0$,
i.e., when the sample is at maximal doping.  In this case, the value of 
$H_c(\beta)$ is the same as in
conventional GL theory with the same $\kappa$.  Thus, far in the normal state,
 the critical field will take the value $H_c(\beta)= (1/\sqrt{2})\sqrt{1 - 
\beta^2}$ since $H_c^0 = 1/\sqrt{2}$ in the conventional SO(2) case 
(see Section~\ref{veconv}).
 
Also of interest is the critical field $H_{c2}$.  
For a type I SC,
at $T<T_c$, when $H > H_c^0$, the SC phase
is destroyed.  After that, if $H$ is reduced, the SC is recovered at
$H_{c2} < H_c^0$ (phenomenon of supercooling).  For a type II SC, $H_{c2}$ 
corresponds to the maximum magnetic field in which a vortex 
lattice can appear; 
it is always higher than $H_c^0$.  Then, $H_{c2}$ is the
strongest field which permits the SC phase to occur. 

Near $H_{c2}$, the order parameter is small (the system is near the
transition), and we can use the linearized SO(5) GL equations. On the
other hand, we can show that, at first order, the microscopic field
$\mathbf{h}$ can be taken equal to the applied field
$\mathbf{H}$.  Beginning with the free energy Eq.~(\ref{free}), 
we choose the Landau gauge $\mathbf{H} = H\hat{z}$ and
$A \hat{y}=Hx$.  Linearizing around $\phi=0$ and $\eta=a_2/\sqrt{b}$
(the values of fields at the transition), the equation for the SC order
parameter takes the form \cite{Levis}:
\begin{displaymath}
\left[ \nabla^2 + \frac{4\pi i}{\Phi_0} \tilde{H} x 
\frac{\partial}{\partial y} - \left( \frac{2 \pi \tilde{H} x}
{\Phi_0}\right)^2  \right] \psi= -\kappa^2 ( 1- \beta) \psi,
\end{displaymath}
where we have set $\tilde{H}=\lambda^2 H$, $\phi = v \psi$ and 
$\Phi_0$ is the flux 
quantum.

This equation can be put in an harmonic oscillator equation form 
for $x$, by seting 
$\psi = f(x) \exp \left( -i( k_y y + k_z z) \right)$ 
and the quantization of the energy leads to a set of allowed values of the 
magnetic field.  
The highest possible value of $H_n$, then, corresponds to the upper critical
field and we find  
\begin{equation}\label{h2}
H_{c2}(\beta) = \frac{\Phi_0}{2 \pi \xi^2}( 1- \beta) 
= \sqrt{2} \kappa H_c^0 ( 1- \beta).
\end{equation}
Note that it is the conventional critical field that appears in the right 
hand of Eq.~(\ref{h2}).

In the GL theory of conventional SC, we have seen that the boundary between
type I and type II SC occurs when $\kappa = \kappa_c = 1/ \sqrt{2}$; 
at this value, $H_{c2}= H_c^0$.
We can already anticipate the value of $\kappa_c$ in the SO(5) model as  
the value for which these two critical fields are equal. 

\begin{figure}[ht]
\centering
\includegraphics[angle=270,width=10cm]{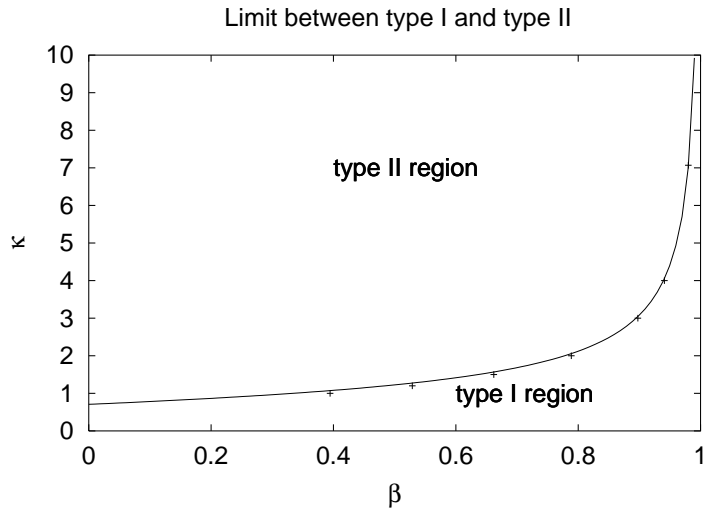}
\caption{$\kappa_c$ as a function of the doping
  $\beta$. 
The curve delimits the different types of SC. 
The points come from numerical calculations (see the end of Section~\ref{SE}
and Table 1).}
\label{k1}
\end{figure}

From (\ref{h1}) and (\ref{h2}) we find: 
\begin{equation}\label{kc}
\kappa_c (\beta) = (1/ \sqrt{2})\sqrt{(1 + \beta)/( 1- \beta)}.
\end{equation}  
Thus we are able to obtain an analytical expression for
$\kappa_c$ as a function of the doping. 
Two facts emerge from Eq.~(\ref{kc}): firstly, we recover (as we must) the 
conventional result when $\beta=0$, and secondly, for any $\beta$, a
value of $\kappa_c$ exists which separates type I and type II behavior
(Figure \ref{k1}). 

As $\beta$ approaches 1 (the SO(5) symmetric case), 
the instability of a vortex 
lattice (in the type I region) will be amplified, and therefore, to reach the 
type II region, we will need a higher $\kappa$. 
In practice, it is easier to vary the doping $\beta$ (or $\mu$) than the GL 
parameter $\kappa$ in a sample.

We can reverse this point of view: if we consider a SC at a fixed value of 
$\kappa$, we will find that, for $\beta>\beta_c$ (defined implicitly
in (\ref{kc})), 
the sample will exhibit type I behavior. 
Thus, for high values of $\kappa$ (i.e., $\kappa\gg1$), 
we also need high $\beta$ (i.e., $\beta$ approaching 1)
in order to see type I behavior. 
We will see in the next section that the numerical calculation of the 
surface energy agrees perfectly with these results.

\subsection{Surface Energy}\label{SE}

In this section, we will derive an expression for the surface energy
in the SO(5) model, and rederive from it numerically the function 
$\kappa_c(\beta)$.  
We will begin by analyzing qualitatively the
surface energy; we will see that for $\kappa$ sufficiently large (small),
the surface energy is negative (positive).  
However, even this qualitative argument indicates that the critical value of
$\kappa$ depends on $\beta$, as anticipated in the previous section.   
We will then present numerical results since, as in the 
conventional model, the surface energy cannot be
evaluated analytically; these numerical results agree with the
conclusion of the previous section based on the critical fields.

\begin{figure}[ht]
\centering
\includegraphics[angle=270,width=7.5cm]{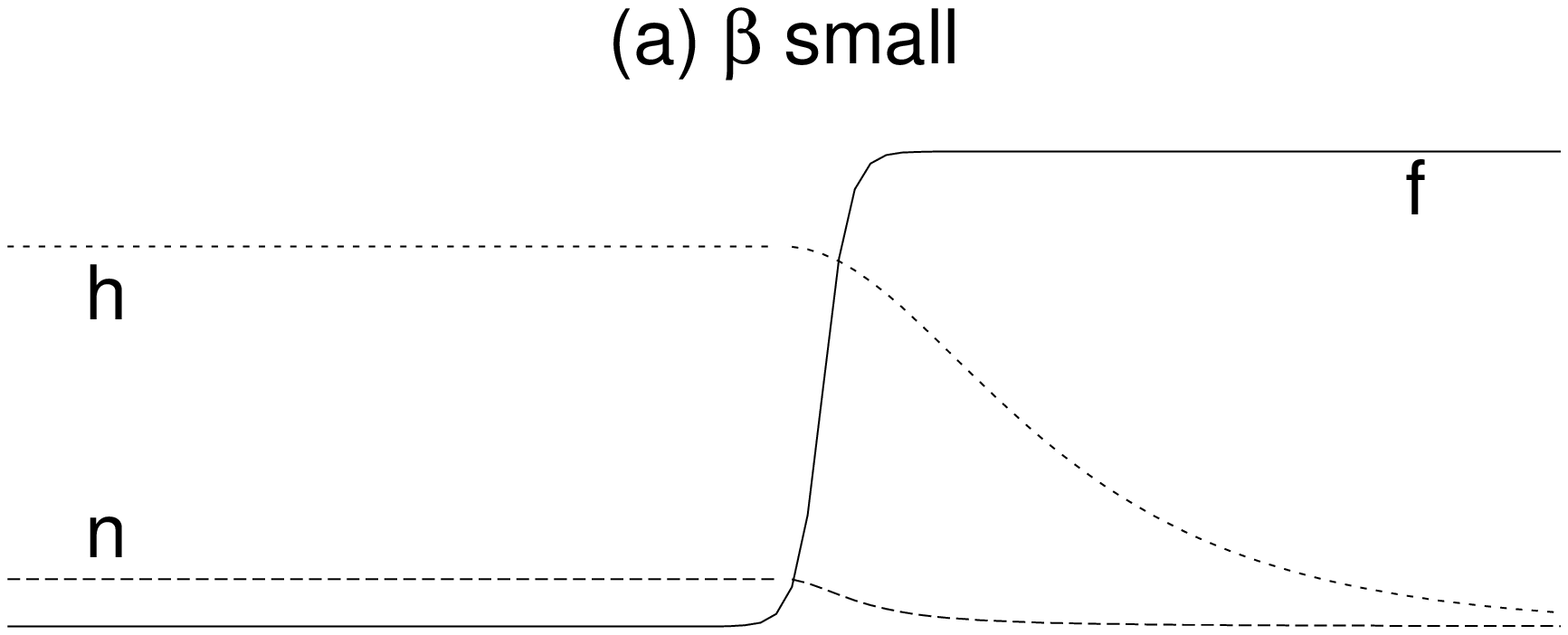}\\ \medskip
\includegraphics[angle=270,width=7.5cm]{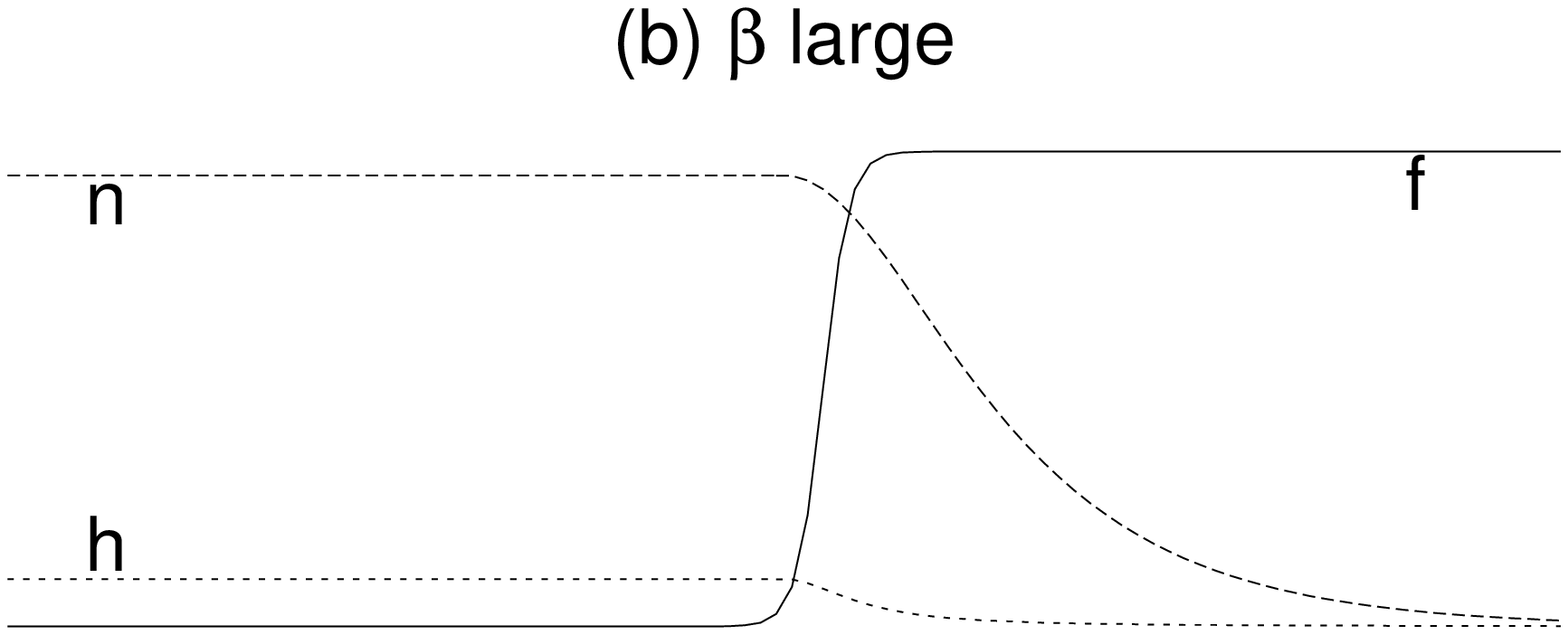}
\caption{Field profiles in surface for 
$\kappa \gg 1/\protect\sqrt{2}$.
(a) $\beta$ small (b) $\beta$ large.}
\label{k2}
\end{figure}

To analyze qualitatively the surface energy,
consider Fig.~\ref{k2}.  In the normal region as $x \to -\infty$, the
positive contribution to the energy coming from the scalar fields $f$
and $n$ exactly cancels the negative condensation energy of the magnetic 
field. In the SC region, as $x \to \infty$, the
scalar field and magnetic energy densities individually are zero.
Inside the surface, variations in the fields contribute to the surface
energy. The details of these variations will determine the sign of the
surface energy.

We can characterize the
fields' variations by their characteristic lengths.  
From Eqs.~(\ref{m1}-\ref{m3}),
these are seen to be
\begin{equation}\label{le}
l_h = 1 , \quad l_f = \frac{1}{\sqrt{2}\kappa}, \quad l_n 
= \frac{1}{ \kappa \sqrt{1- \beta}}  .
\end{equation}

At issue is whether in going from one asymptotic regime to the other,
the scalar fields or the magnetic field are capable of changing
more quickly. If the scalar fields vary faster, the magnetic field will
penetrate into the superconducting region, and the negative
magnetic energy density in this transition zone will not be
accompanied by a positive scalar energy density, resulting in a
negative surface energy. If
the magnetic field varies faster, the situation is reversed and the
surface energy will be positive.
Thus we need to compare the larger of $l_n$ and $l_f$ with $l_h$.
We see from Eq.~(\ref{le})
that $l_n$ is always greater than $l_f$, so $l_n$ determines how
quickly the scalar fields can vary.
Two cases are important to us when $ \kappa
\gg 1/ \sqrt{2}$.  First, when $\beta$ is small (Fig.~\ref{k2}a), we find from
(\ref{le}) that $l_h>l_n$, and as described above, the surface energy
will be negative.  This corresponds to the behavior of a type II SC.

On the other hand, when $\beta$ is large (Fig.~\ref{k2}b), we have the 
opposite situation: $l_h<l_n$, and the surface energy will be positive.
This is the case of a type I SC. 

By continuity, for any $\kappa$, there must exist a $\beta$ where the surface 
energy is zero.
This provides an alternative way of finding $\kappa(\beta)$, and must agree 
with Eq.~(\ref{kc}).

We define the SO(5) surface energy in the same way as in Section 
\ref{seconv}.  
Subtracting Eq.~(\ref{gn}) from Eq.~(\ref{legendre})
we are led to
\begin{equation}
 \hat{G}_s =\int_{-\infty}^{\infty}{dx} \left\{ f_s 
- \frac{\hat{h} H_c}{4 \pi} - \left( f_n - \frac{a_2^4}{4b} 
- \frac{H_c^2}{8 \pi} \right) \right\}.
\end{equation}
Using (\ref{fs}) we find, after some algebra,
\[
\hat{G}_s = \frac{ (H_c^0)^2}{4 \pi} \lambda I,
\] 
where
\begin{equation}\label{I}
 I = \int_{-\infty}^{\infty}{ds} \left\{ \frac{1}{2}(1-f^4 - n^4) 
- f^2n^2 + h^2 - \sqrt{2}h\sqrt{ 1 - \beta^2} \right\},
\end{equation}
with $\lambda$ and $H_c^0$ defined in Section~\ref{sc}. 
Therefore, the $\beta$ 
dependence appears only in $I$, and the expression reduces to the 
conventional one, Eq.~(\ref{notreamii}), when we set $n= \beta =0$.

Of interest are the conditions under which this integral is zero. It will be 
recalled that in the SO(2) case, when $\kappa = 1/ \sqrt{2}$, not only is the 
integral zero but the integrand is zero -- a stronger statement, 
and one which is 
explained by Bogomol'nyi \cite{Bogol}. One might hope that the same feature 
would be found in the SO(5) model, along with an analogous 
analytical explanation.
Unfortunately, such is not the case, and the surface energy 
must be studied numerically to find $\kappa_c( \beta )$.

\begin{figure}[ht]
\centering
\includegraphics[angle=270, width=7.5cm]{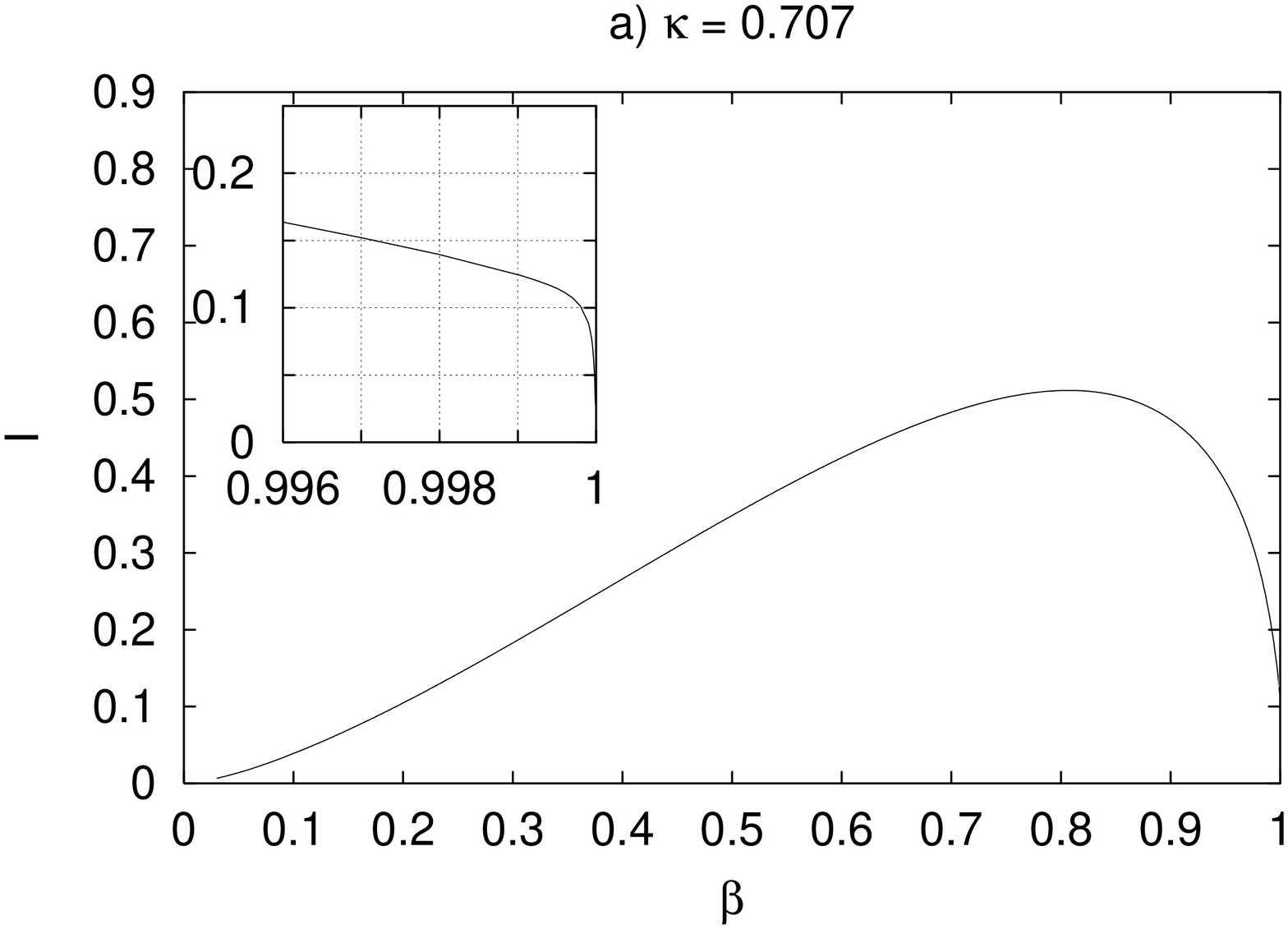}\\ \medskip
\includegraphics[angle=270, width=7.5cm]{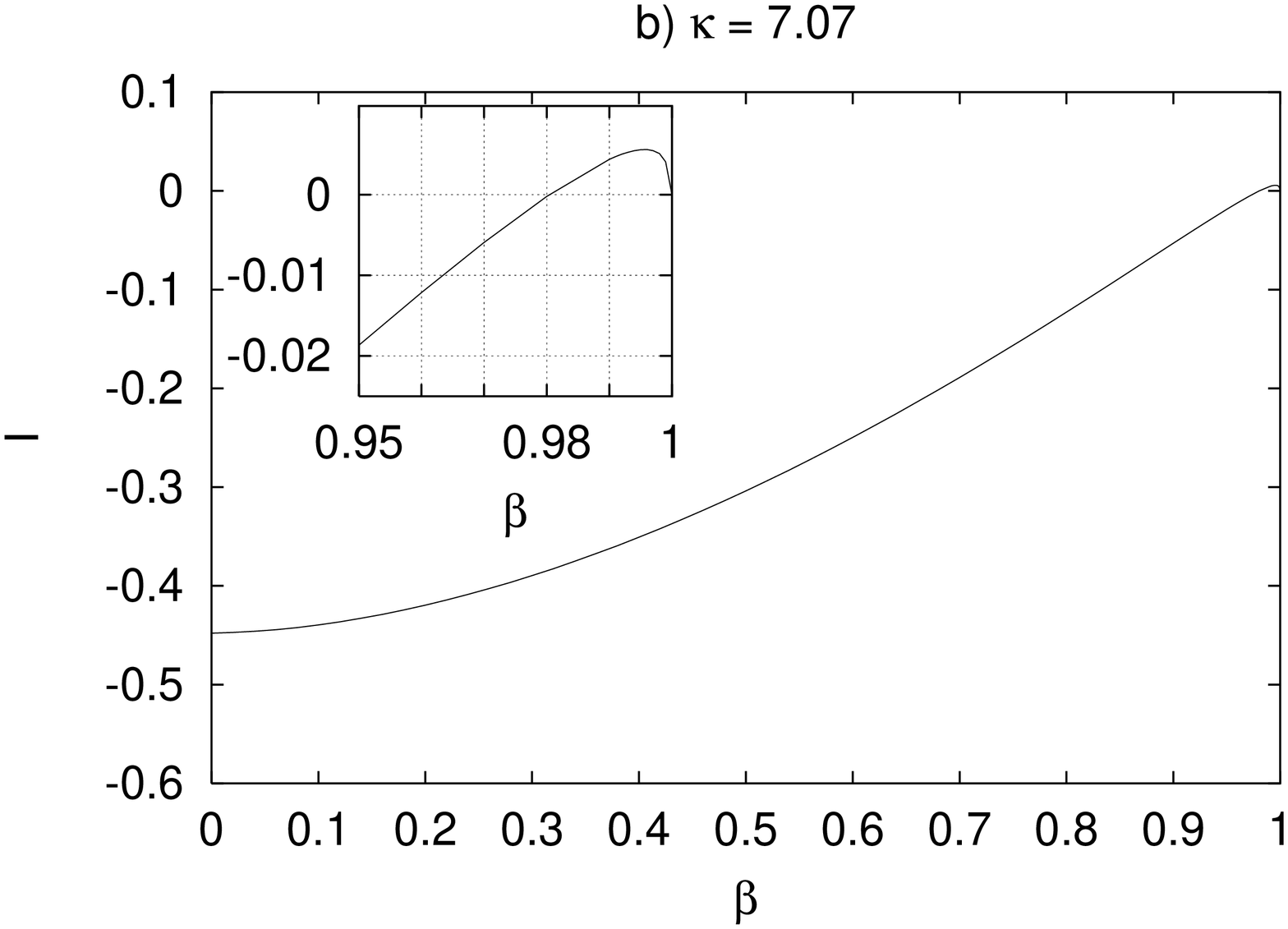}
\caption{Surface energy $I(\beta)$ for
a) $\kappa=0.707$ and b) $\kappa=7.07$.}
\label{k4}
\end{figure}

We have solved Eqs.~(\ref{m1}-\ref{m3}) numerically, using a
relaxation algorithm. The boundary conditions appropriate for the
problem
are:
\begin{equation}\label{bond}
\begin{array}{cccc}
x \rightarrow -\infty : & \quad f \rightarrow 0, & 
\quad h \rightarrow \sqrt{1- \beta^2}/\sqrt{2}, 
& \quad n \rightarrow \sqrt{\beta};  \\
x \rightarrow \infty : &  f \rightarrow 1, 
& h \rightarrow 0, &  n \rightarrow 0 .
\end{array}
\end{equation}
Having found the solution, it is easy to compute the dimensionless
surface energy $I$ as a function of the doping $\beta$ by numerical
integration of (\ref{I}).  We present, in Fig.~\ref{k4}, the results for
$\kappa=0.707$ and $\kappa=7.07$ respectively.  The case $\kappa <
1/\sqrt{2}$ is of no interest since the surface energy is
positive for all $\beta$. 

Fig.~\ref{k4}a corresponds to $\kappa_c$ for
conventional SC.  First of all, we see that the surface energy is zero for
$\beta=0$, as expected since, at this particular value, we 
recover the SO(2) case.  For all other values of $\beta$, we have a positive 
surface energy. Thus, the sample remains in the type I region 
for any value of the doping. This is in agreement 
with Eq.~(\ref{kc}), as seen in Fig.~\ref{k1}. Note also that the energy 
reaches a maximum and decreases until it reaches zero again at $\beta=1$. 
At this point, $h=0$ throughout the sample (see Eq.~(\ref{bond})), and the 
ground state is SO(5) symmetric.  Hence, no broken symmetry appears and for
topological reasons (since the surface energy can be treated like a
soliton), the surface energy must be zero.

A more interesting case is presented in Fig.~\ref{k4}b, corresponding to 
$\kappa =7.07$ 
(in the conventional case, a type II superconductor).  The curve starts at 
$\beta=0$ at the conventional value and passes through zero at $\beta 
\approx 0.98$, in agreement with the analytical result (\ref{kc}). Below this 
critical value, the sample has type II 
behavior since the surface energy is negative.  
Above, the surface energy is positive, corresponding to type I behavior, 
i.e., vortices in this region are not energetically stable. 
Observe that the surface energy is again zero for $\beta=1$.  

We can determine in this way $\kappa_c$ as a function of $\beta$; 
the results are reported in Table 1, and are in excellent 
agreement with Eq.~(\ref{kc}).

\begin{center}
\begin{tabular}{|c|c|c|}
\hline
$\beta$ & $\kappa_c(\beta)$ (numerical) & From (\ref{kc}) \\
\hline \hline
0.3945 & 1 & 1.07\\
0.5288 & 1.2 & 1.27\\
0.6620 & 1.5 & 1.57\\
0.7887 & 2  & 2.06\\
0.8976 & 3  & 3.04\\
0.9410 & 4  & 4.06\\
0.98   & 7.07 & 7.04\\
0.9998 & 70.7 & 70.71\\
\hline
\end{tabular}
\end{center}
\begin{center}
Table 1: Comparison of
$\kappa_c$ obtained numerically via vortex energetics and surface
energy, and analytically via Eq.~(\ref{kc}).
\end{center}

\subsection{Vortex Energetics in the SO(5) Model}

As discussed in Section~\ref{veconv}, the magnetic behavior of a
superconductor can also be determined by studying vortex energetics
\cite{Mack}.  Thus, one can calculate the free energy per quantum of
magnetic flux of an $m$-vortex $\FF(m)=F(m)/m$; if this function is of
positive or negative slope, the superconductor is of type II or I,
respectively; zero slope indicates the critical case, which can be
expressed in terms of $\kappa_c(\beta)$.  In the SO(5) model, the
numerical calculation of $\FF(m)$ for $\kappa=7.07$ and $70.7$ are
given, respectively, in Fig.~2c and Fig.~2d in \cite{Mack}. From the
first of these, we can see that for $\beta < 0.98$ $\FF(m)$ is of
positive slope, while for $\beta > 0.98$ it is of negative slope; to a
very good approximation, the slope is zero at $\beta=0.98$. Thus,
$\kappa_c(\beta=0.98)=7.07$. Similarly, Fig.~2d of the same reference
indicates $\kappa_c(\beta=0.9998)=70.7$. A comparison of this method
with the analytical method discussed in Section~\ref{hso5}
is shown in Table
1; the agreement is excellent.

\section{Summary and Conclusions}

In this paper, magnetic properties of superconductors were discussed,
with an eye towards analyzing the SO(5) model of high-temperature
superconductivity.  The Ginzburg-Landau theory of conventional
superconductivity was discussed in Section~\ref{sc}, mainly to review some
fundamental and well-known results and also to establish notation. The
energy density of a surface separating superconducting and normal
regions at critical applied field, and its importance for determining
the behavior of the superconductor, were discussed.  In addition, we
reviewed the approach of Bogomol'nyi wherein the vortex energy is
written in a form that shows clearly that at $\kappa_c$, the energy is
proportional to the winding number of the vortex.  Finally, we
discussed an alternative approach to determining the critical field
which is suggested by Bogomol'nyi's work, namely, studying the energy
of an $m$-vortex as a function of its winding number. Both
analytically and numerically, this approach
can be shown to be equivalent to the surface energy approach.

In the second part, we introduced briefly the SO(5) model, and
derived, in the context of that model, expressions for the critical
magnetic fields, $H_c(\beta)$ and $H_{c2}(\beta)$. These enable us to
calculate an analytic expression for $\kappa_c(\beta)$. Next, we
studied the surface energy of a boundary between superconducting and
non-superconducting regions in an applied critical field. Finally, we
discussed briefly an alternative approach based on vortex energetics.

It is worth comparing the two numerical approaches to calculating
$\kappa_c$, namely, by calculating the surface energy or by vortex
energetics. In the former approach, one must solve the field equations
(for instance, (\ref{m1}-\ref{m3}) in the SO(5) model), evaluate the
surface integral (\ref{I}), and vary $\beta$ for fixed $\kappa$ (or
vice-versa) until the surface integral vanishes. In the latter
approach, one must solve the vortex field equations (the
generalization of (\ref{motion3}) and (\ref{motion4}) to SO(5)), evaluate
the energy as a function of winding number, and determine for which
$\kappa,\beta$ the function $\FF(m)$ is independent of $m$ for large
$m$. The latter method is somewhat more demanding numerically,
essentially since several values of $m$ must be considered for each
$\kappa,\beta$, though
it is perhaps more intuitive, since one can think of type I behavior
as due to an energetic preference for the coalescence of many single
vortices into one large $m$-vortex.

The main new result in this paper is Eq.~(\ref{kc}), which
expresses the value of $\kappa$ as a function of $\beta$ corresponding
to the boundary between type I and type II superconductors.  This
result is rather important: it states that as $\beta\to 1$ (i.e., as
the doping is reduced to its critical value), then even with a large
value of $\kappa$, the material would be a type I superconductor,
contrary to what one would naively expect. Some implications of
this have been discussed in \cite{Mack}.

\end{document}